\documentclass[11pt]{article}
\usepackage[applemac]{inputenc}

\usepackage{epsf}
\usepackage{color}
\usepackage{epsfig}

\usepackage{graphicx}
\usepackage{amsmath}
\usepackage{latexsym}
\usepackage{amssymb}

\usepackage{alltt}

\usepackage{a4}

\newtheorem{theorem}{Theorem}[section]

\numberwithin{equation}{section}

\newcommand{\ad}{{\rm ad}}
\newcommand{\spn}[1]{{\rm span}(#1)}
\newcommand{\e}{{\rm e}}
\newcommand{\R}{R}

\newcommand{\N}{\ensuremath{\mathbb{N}}}

\newcommand{\I}{\ensuremath{\mathcal{I}}}
\newcommand{\J}{\ensuremath{\mathcal{J}}}

\begin{document}

\title{Efficient computation of the Zassenhaus formula}

\author{Fernando Casas$^{1}$\thanks{Email: \texttt{Fernando.Casas@uji.es}}
   \and
Ander Murua$^{2}$\thanks{Email: \texttt{Ander.Murua@ehu.es}}
   \and
  Mladen Nadinic$^{3}$ 
   }

%\date{}
\maketitle

\begin{abstract}

A new recursive procedure to compute the Zassenhaus formula up to high order is presented, 
providing each exponent in the factorization directly as a linear combination of independent 
commutators and thus
containing the minimum number of terms. The recursion can be easily implemented in a symbolic
algebra package and requires much less computational effort, both in time and memory resources, than
previous algorithms. In addition, by  bounding appropriately each term in the recursion, it is possible
to get a larger convergence domain of the Zassenhaus formula when it is formulated in a Banach
algebra.

\vspace*{1cm}

\begin{description}
 \item $^1$Institut de Matem\`atiques i Aplicacions de Castell\'o and 
   Departament de Matem\`atiques, Universitat Jaume I,
  E-12071 Castell\'on, Spain.
 \item $^2$Konputazio Zientziak eta A.A. saila, Informatika
Fakultatea, EHU/UPV, Donostia/San Sebasti\'an, Spain.
 \item $^3$ Departamento de Matem\'atica, Facultad de Ciencias, Universidad de Santiago
 de Chile, Avda. Bernardo O'Higgins 3363, Estaci\'on Central, Santiago, Chile.
\end{description}

\end{abstract}

\section{Introduction}   \label{sec.1}

Products of exponentials of non-commuting indeterminate variables are of fundamental importance in physics and
mathematics. As is well known, the Baker--Campbell--Hausdorff theorem states that $\e^X \e^Y = \e^Z$,
with
\begin{equation}    \label{eq.1}
    Z = \log ( \e^X \e^Y) = X + Y + \sum_{m=2}^{\infty} Z_m (X,Y).
\end{equation}
Here $Z_m(X,Y)$ is a homogeneous Lie polynomial in the non-commuting variables $X$ and $Y$. In other
words, $Z_m$ is a linear combination (with rational coefficients) of commutators of the form
$[V_1,[V_2, \ldots,[V_{m-1},V_m]
\ldots]]$ with $V_i \in \{X,Y\}$ for $1 \le i \le m$. We recall that $[X,Y] \equiv  X Y - Y X$.
The first terms in the series (\ref{eq.1}) read explicitly
 \[
   Z_2  =  \frac{1}{2} [X,Y], \quad
   Z_3  =  \frac{1}{12} [X,[X,Y]] - \frac{1}{12}
                 [Y,[X,Y]], \quad 
   Z_4  =  \frac{1}{24} [X,[Y,[Y,X]]].  
\]
The expression $\e^X \, \e^Y = \e^Z$ is then properly called  the \emph{Baker--Campbell--Haus\-dorff
formula}  (BCH for short) and plays a fundamental role in many fields of mathematics
(theory of linear differential
equations \cite{magnus54ote}, Lie groups  \cite{gorbatsevich97fol},
numerical analysis \cite{hairer06gni}), theoretical physics
(perturbation theory \cite{dragt76lsa}, Quantum Mechanics \cite{weiss62tbf},
Statistical Mechanics \cite{kumar65oet,wilcox67eoa}, quantum computing 
\cite{sornborger99hom}), control theory (analysis and design of nonlinear
control laws, nonlinear filters, stabilization of rigid bodies \cite{torres05asp}), etc. (see \cite{bonfiglioli} for
a comprehensive treatment of the algebraic aspects of the BCH formula).

Although the BCH theorem establishes the precise algebraic structure of the exponent $Z$ in (\ref{eq.1}),
it does not provide simple ways to compute explicitly this series. As a matter of fact, the problem of
effectively computing the BCH series up to arbitrary degree has a long history, and different procedures have 
been proposed along the years, starting with the work of Richtmyer and Greenspan in 1965 (see 
\cite{casas09aea} for a review). Most of
the procedures lead to expressions where not all the iterated commutators are linearly independent
(due to the Jacobi identity and other identities appearing at higher degrees). Equivalently, the resulting
expressions are not formulated directly in terms of a basis of the free Lie algebra $\mathcal{L}(X,Y)$
generated by the symbols $X$ and $Y$. Of course, it is always possible to write these expressions in
terms of a basis, but this rewriting process is time consuming and require considerable memory
resources. This in addition is made more difficult due to the exponential growth of the number of terms
with the degree $m$. Recently, a new efficient algorithm has been proposed which allows one to get closed
expressions for $Z_m$ up to a very high degree in terms of both the classical Hall basis and 
the Lyndon basis of $\mathcal{L}(X,Y)$ \cite{casas09aea}.

In the paper dealing with the expansion bearing his name, Magnus \cite{magnus54ote} 
cites an unpublished reference by Zassenhaus, reporting that 
there exists a formula which may be called the dual of the (Baker--Campbell--)Hausdorff formula. This result
can be stated as follows.

\begin{theorem}
  (Zassenhaus Formula). Let $\mathcal{L}(X,Y)$ be the free Lie algebra generated by $X$ and $Y$. Then, $\e^{X + Y}$ can be uniquely decomposed  as
  \begin{equation}  \label{zass.1}
      \e^{X + Y} = \e^X \, \e^Y \, \prod_{n=2}^{\infty} \e^{C_n(X,Y)} = \e^X \, \e^Y \, \e^{C_2(X,Y)} \, 
      \e^{C_3(X,Y)} \, \cdots  \, \e^{C_n(X,Y)} \, \cdots,
  \end{equation}    
  where $C_n(X,Y) \in  \mathcal{L}(X,Y)$ is a homogeneous Lie polynomial in $X$ and $Y$ of degree $n$.
\end{theorem}      

The existence of such a  formula is an immediate consequence of the BCH theorem. In fact, it is clear that
$\e^{-X} \e^{X+Y} = \e^{Y + D}$, where $D$ involves Lie polynomials of degree $> 1$. Now
$\e^{-Y} \e^{Y + D} = \e^{C_2 + \tilde{D}}$, where $\tilde{D}$ involves Lie polynomials of degree $> 2$ and 
the process is repeated again. Induction allows one to get the general result.

By comparing with the BCH formula it is possible to obtain the first terms of the formula (\ref{zass.1}) as
  \[
     C_2(X,Y) = -\frac{1}{2} [X,Y], \qquad\quad  C_3(X,Y) = \frac{1}{3} [Y,[X,Y]] + \frac{1}{6} [X,[X,Y]].
  \]   

Although less familiar than the BCH formula, the Zassenhaus formula constitutes nevertheless
a standard tool in several fields, since the disentangling of the exponential of the sum of two non-commuting
operators into an (in general infinite) product of exponential operators arises for instance in
statistical mechanics, many-body theories, quantum optics, path integrals, $q$-analysis in
quantum groups, etc. \cite{quesne04dqe}. Also in particle accelerators physics, Dragt and his collaborators
have used the Zassenhaus formula to compute the relevant maps both in Taylor and factored product form
\cite{dragt12lmf}. Yet in another context, very recently Iserles and Kropielnicka 
\cite{iserles11eaf} have proposed a
new family of high-order splitting methods for the numerical integration of the time-dependent
Schr\"odinger equation based on a symmetric version of the Zassenhaus formula.

Several systematic computations of the terms $C_n$ for $n >3$ in the 
Zassenhaus formula have been carried out
in the literature, starting with the work of Wilcox \cite{wilcox67eoa}, 
where a recursive procedure is presented that has been
subsequently used to get explicit expressions up to $C_6$ in terms of nested commutators 
\cite{quesne04dqe}. On
the other hand, Volkin \cite{volkin68ica} proposed a general technique to express a function
of non-commuting operators as an expansion in successively higher commutators of the operators involved.
In particular, he was able to get recursive formulae up to $C_9$. By following an idea already suggested by
Wilcox in \cite{wilcox67eoa}, Suzuki \cite{suzuki77otc}
obtained the successive terms $C_n(X,Y)$ in
\begin{equation}   \label{zass.2}
  \e^{\lambda(X+Y)} = \e^{\lambda X} \, \e^{\lambda Y} \, 
   \e^{\lambda^2 C_2} \, \e^{\lambda^3 C_3} \cdots
\end{equation}   
by differentiating both sides with respect to $\lambda$ and setting $\lambda = 0$
after each differentiation. In this way
\begin{eqnarray*}
    C_2 & = & \frac{1}{2} \left( \frac{d^2 }{d \lambda^2} ( \e^{-\lambda Y} \e^{-\lambda X}
         \e^{\lambda(X+Y)} ) \right)_{\lambda=0}  = \frac{1}{2} [Y,X]  \\
    C_3 & = & \frac{1}{3!} \left( \frac{d^3 }{d \lambda^3} ( \e^{-\lambda^2 C_2} \e^{-\lambda Y} 
          \e^{-\lambda X}
         \e^{\lambda(X+Y)} ) \right)_{\lambda=0}  = \frac{1}{3} [C_2,X + 2 Y]  
\end{eqnarray*}
and in general, for $n \ge 3$,
\begin{equation}   \label{suzu.1}         
      C_n = \frac{1}{n!} \left( \frac{d^n }{d \lambda^n} (  \e^{-\lambda^{n-1} C_{n-1}}
      \cdots \e^{-\lambda^2 C_2} \e^{-\lambda Y} \e^{-\lambda X}
         \e^{\lambda(X+Y)} ) \right)_{\lambda=0}. 
 \end{equation}
Finally, Baues \cite{baues81cca} gave explicit
formulae for the Zassenhaus terms via homotopy theory and more recently Kurlin \cite{kurlin}
obtained a closed expression for $\sum_{n\ge 2} C_n$ in the metabelian case. 

All of these proposals give results for $C_n$ as a linear combination of nested commutators. 
In contrast,
Scholz and Weyrauch \cite{scholz06ano} proposed a recursive procedure
based on upper triangular matrices that can be easily implemented in a symbolic algebra package. In this
case, however, the expressions for $C_n$ are not explicitly written down in terms of homogeneous Lie
polynomials. More recently \cite{weyrauch09ctb}, the same authors have applied
 a technique related to one previously used by 
Witschel \cite{witschel75ooe}
to get $C_n$ up to $n=15$ in less than 2 minutes of CPU time. Here also the Zassenhaus terms are written 
as 
\begin{equation}  \label{ne.2}
   C_n =  \sum_{w, |w|=n} \,
     g_w \, w,
\end{equation}
where $g_w$
is a rational coefficient and the sum is taken over all words
$w$ with length $|w|=n$ in the symbols $X$ and $Y$, i.e., $w = a_1 a_2 \cdots a_n$,
each $a_i$ being $X$ or $Y$. Of course, one may always apply the Dynkin--Specht--Wever theorem
\cite{jacobson79lal}, and express $C_n$ as 
\begin{equation}   \label{eq.2.2}
  C_n =  \, \frac{1}{n} \, \sum_{w, |w|=n} \,
     g_w \, [w],
\end{equation}
that is, the individual terms are the same as in the associative
series (\ref{ne.2}) except that the word $w= a_1 a_2 \ldots a_n$
is replaced with the right nested
commutator $[w] = [a_1,[a_2, \ldots [a_{n-1},a_n]\ldots]]$ and
the coefficient $g_w$ is divided by the word length $n$. In this way $C_n$ is constructed as
a linear combination of nested commutators of homogeneous degree $n$, that is, as a linear combination of elements of the homogeneous subspace $\mathcal{L}(X,Y)_n$ of degree $n$ of the free Lie algebra  
$\mathcal{L}(X,Y)$. As a matter of fact, another representation of (\ref{ne.2})
in terms of nested commutators 
is proposed in \cite{weyrauch09ctb} which, it is claimed, contains less terms than the Dynkin--Specht--Wever
representation. In any case, 
 it should be stressed that the set of nested commutators $[w]$ 
for words $w$ of length $n$ in either representation  
is \emph{not} a basis of the homogeneous subspace $\mathcal{L}(X,Y)_n$.

The purpose of this paper is twofold. First, to present a new recurrence 
that allows one to express the Zassenhaus terms $C_n$ directly as a linear combination of independent
elements of the homogeneous subspace  $\mathcal{L}(X,Y)_n$. In other words, the procedure, which
can be easily implemented in a symbolic algebra package, gives $C_n$ 
up to a prescribed degree  directly in terms of independent commutators involving $n$ operators $X$ and
$Y$. In this way, no rewriting process in a %Hall--Viennot 
basis of  $\mathcal{L}(X,Y)$ is necessary, thus
saving considerable computing time and memory resources. Moreover, we are able to express directly
$C_n$ with the minimum number of commutators required at each degree $n$.

The second aspect we are dealing with concerns the convergence of the Zassenhaus formula
when it is formulated in a Banach algebra. As
far as we know, there are only two previous results in the literature. 
The first one was obtained by Suzuki \cite{suzuki77otc} starting with the recurrence
(\ref{suzu.1}). Specifically, he was able to prove that if $|\lambda| (\|X\| + \|Y\|) \le \log 2 - 1/2$ the infinite product
(\ref{zass.2}) converges.  Subsequently, Bayen \cite{bayen79otc} generalized the analysis, showing
that the product (\ref{zass.2}) converges if $|\lambda| (\|X\| + \|Y\|) \le r$, where $r\approx 0.596705$
is a root of a certain transcendental equation. In the present work, % by following Bayen's methodology, 
we obtain sharper bounds for the terms of the Zassenhaus formula which show that the product (\ref{zass.1}) converges in an enlarged domain.

A simple but important remark is in order here. In some applications,  
the ``left-oriented''  Zassenhaus formula 
\begin{equation}   \label{zass.1.1}
    \e^{X + Y} = \cdots \, \e^{\hat{C}_4(X,Y)} \, \e^{\hat{C}_3(X,Y)} \, \e^{\hat{C}_2(X,Y)} \,
        \e^Y \, \e^X
\end{equation}        
is used instead of (\ref{zass.1}). A simple observation shows that the exponents $\hat{C}_i$ and $C_i$
are related through
\[
    \hat{C}_i(X,Y) = (-1)^{i+1} C_i(X,Y), \qquad i \ge 2
\]    
and so we may restrict ourselves to analyzing the ``right-oriented''  formula  (\ref{zass.1}).

\section{Constructing the Zassenhaus terms}

\subsection{A new recurrence}

To derive our recursive procedure, it is convenient to introduce a parameter $\lambda$ as in
(\ref{zass.2}), 
\begin{equation}   \label{zass.2.1}
  \e^{\lambda(X+Y)} = \e^{\lambda X} \, \e^{\lambda Y} \, 
   \e^{\lambda^2 C_2} \, \e^{\lambda^3 C_3} \, \e^{\lambda^4 C_4} \cdots
\end{equation}   
so that the original Zassenhaus formula (\ref{zass.1}) is recovered when $\lambda=1$. Moreover, we
consider the compositions
\begin{equation}   \label{rec.1}
   \R_1(\lambda)=\e^{-\lambda Y} \, \e^{-\lambda X} \, \e^{\lambda(X+Y)}
\end{equation}
and  for each $n\geq 2$,
\begin{equation}
  \label{eq:R_n}
\R_n(\lambda) =  \e^{-\lambda^{n} C_{n}}
      \cdots \, \e^{-\lambda^2 C_2} \, \e^{-\lambda Y} \, \e^{-\lambda X} \, 
         \e^{\lambda(X+Y)} = \e^{-\lambda^{n} C_{n}} \, \R_{n-1}(\lambda).
\end{equation}
It is then clear that  
\begin{equation}
  \label{eq:R_n2}
  \R_n(\lambda) =  \e^{\lambda^{n+1} C_{n+1}}  \, \e^{\lambda^{n+2} C_{n+2}} \cdots
\end{equation}
Finally, we introduce
  \begin{equation}
\label{eq:F_nDef}
    F_n(\lambda) \equiv  \left(
\frac{d}{d \lambda} \R_n(\lambda) 
\right) \R_n(\lambda)^{-1}, \qquad n \ge 1.
  \end{equation}
To determine the series $F_n(\lambda)$ 
we proceed as follows. 
On the one hand, a simple calculation starting from (\ref{eq:R_n}), leads for $n\geq 2$ to
\begin{eqnarray}  \label{eq:F_n.n}
  F_n(\lambda) &=&  -n \, C_n \, \lambda^{n-1}
+ \e^{-\lambda^n C_n} \, \left( \frac{d }{d \lambda} R_{n-1}(\lambda) \right) R_{n-1}(\lambda)^{-1} \,  \e^{\lambda^n C_n}  \nonumber  \\
   & = &   -n \, C_n \, \lambda^{n-1}
+ \e^{-\lambda^n C_n} \, F_{n-1}(\lambda) \, \e^{\lambda^n C_n}  \nonumber \\
&=&  -n \, C_n \, \lambda^{n-1}+ \e^{-\lambda^n \ad_{C_n}} F_{n-1}(\lambda)   \nonumber \\
&=&  \e^{-\lambda^n \ad_{C_n}} (F_{n-1}(\lambda) -n \, C_n \, \lambda^{n-1}),
\end{eqnarray}
where we have used the well known formula
\[
    \e^{A} B \e^{-A} = \e^{\ad_A} B = \sum_{n\ge 0} \frac{1}{n!} \ad_A^n B
\]
with
\[
\ad_A B = [A,B], \qquad  \ad_A^j B = [A, \ad_A^{j-1} B], \qquad \ad_A^0 B = B.
\]

On the other hand, differentiating expression (\ref{eq:R_n2}) with respect to $\lambda$  and taking
into account (\ref{eq:F_nDef}) we arrive at
\begin{equation}   \label{rec.n2}
  F_n(\lambda) = (n+1) \, C_{n+1} \, \lambda^{n} + \sum_{j=n+2}^{\infty} j \, \lambda^{j-1} \,
  \e^{\lambda^{n+1} \ad_{C_{n+1}}} \cdots \e^{\lambda^{j-1} \ad_{C_{j-1}}} C_j.
\end{equation} 
In other words,
\begin{equation*}
  F_n(\lambda) = (n+1) C_{n+1} \, \lambda^{n} + G_{n+1}(\lambda), \qquad n \ge 1,
\end{equation*}
where $G_{n+1}(0)=G_{n+1}^{(1)}(0)=\cdots = G_{n+1}^{(n)}(0)=0$. In consequence, we have, for
$n\geq 1$,
\begin{eqnarray}
 F_{n+1}(\lambda) &=& 
\label{eq:FG}
\e^{-\lambda^{n+1} \ad_{C_{n+1}}} \, G_{n+1}(\lambda), \\
C_{n+1} &=& \frac{1}{(n+1)!} \, F_n^{(n)}(0), \\
\label{eq:G_n}
G_{n+1}(\lambda) &=& F_n(\lambda) - \frac{\lambda^n}{n!} F_n^{(n)}(0).
\end{eqnarray}
Expressions (\ref{eq:FG})--(\ref{eq:G_n}) allow one to compute recursively the Zassenhaus terms $C_n$
 starting from $F_1(\lambda)$. The sequence is 
 \[
   F_n(\lambda) \;\; \longrightarrow \;\; C_{n+1} \;\; \longrightarrow \;\; G_{n+1}(\lambda)
   \;\; \longrightarrow \;\; F_{n+1}(\lambda) \;\; \longrightarrow \;\; \cdots, \quad n \ge 1.
\]
Let us analyze more in detail this procedure, with the goal of providing an algorithm well adapted from
a computational point of view.

For $n=1$, and taking into account (\ref{rec.1}), we get
\begin{eqnarray*}
 F_1(\lambda) 
&=& 
-Y -\e^{-\lambda \, Y} X \e^{\lambda \, Y} 
+  \e^{-\lambda \, Y}  \e^{-\lambda \, X} (X+Y)  \e^{\lambda \, X}  \e^{\lambda \, Y} \\
&=& 
-Y - \e^{-\lambda \ad_{Y}} X +  \e^{-\lambda \ad_{Y}}  \e^{-\lambda \ad_{X}} (X+Y)\\
&=&
\e^{-\lambda \ad_{Y}}  (\e^{-\lambda \ad_{X}}-I) Y,
\end{eqnarray*}
that is,
\begin{eqnarray}
  \label{eq:F_1}
 F_1(\lambda) &=& \sum_{i=0}^{\infty}\sum_{j=1}^{\infty} \frac{(-\lambda)^{i+j}}{i!j!} \ad_{Y}^{i}\ad_{X}^j Y
\end{eqnarray}
or equivalently
 \begin{equation}   \label{eq:F_1.2}
     F_1(\lambda) =  \sum_{k=1}^{\infty} f_{1,k} \, \lambda^k, \qquad \mbox{ with } \qquad 
    f_{1,k} =  \sum_{j=1}^k \frac{(-1)^k}{j! (k-j)!} \ad_Y^{k-j} \ad_X^j Y.
 \end{equation}
In general, from (\ref{eq:F_n.n}) a straightforward calculation  shows that for $n \ge 2$,
   \begin{equation}   \label{rec.2.3}
     F_n(\lambda) = \sum_{k=n}^{\infty}  f_{n,k} \, \lambda^k, \qquad \mbox{ with } \qquad 
        f_{n,k} = \sum_{j=0}^{[k/n]-1}  \frac{(-1)^j}{j!} \ad_{C_n}^{j} f_{n-1,k-nj}, \qquad\quad  k \ge n.
 \end{equation}
Here $[k/n]$ denotes the integer part of $k/n$. Moreover, a closer examination of (\ref{rec.n2}) reveals that
\begin{eqnarray}  \label{rec.2.4}
  F_n(\lambda) &=& (n+1) C_{n+1} \lambda^{n} + 
(n+2) \e^{\lambda^{n+1}\ad_{C_{n+1}}} C_{n+2} \lambda^{n+1}  + \cdots   \nonumber \\
& = & (n+1) C_{n+1} \lambda^{n} + (n+2) C_{n+2} \lambda^{n+1} + \cdots   \nonumber \\
& &+  (2n+2) C_{2n+2} \lambda^{2n+1}+ \lambda^{2n+2} [C_{n+1},C_{n+2}] + \cdots  \nonumber \\
&=& \sum_{k=n+1}^{2n+2} k \, C_{k} \, \lambda^{k-1} + \lambda^{2n+2} H_n(\lambda),
\end{eqnarray}
where $H_n(\lambda)$ involves commutators of $C_j$, $j \ge n+1$. Notice that
the terms $C_{n+1}, \ldots,C_{2n+2}$ of the Zassenhaus formula can be then directly 
obtained from $F_{n}(\lambda)$.
In particular, one directly gets from (\ref{eq:F_1}) 
\begin{eqnarray}   \label{c.prim}
  C_{n+1} = \frac{1}{n+1} f_{1,n} = \frac{1}{n+1} \sum_{i=0}^{n-1} \frac{(-1)^{n}}{i!(n-j)!} \ad_{Y}^{i}\ad_{X}^{n-j} Y, \quad 
\mbox{for} \quad n=1,2,3.
\end{eqnarray}
Explicitly,
\begin{eqnarray*}
C_2 &=& -\frac{1}{2}\, [X,Y], \\ 
C_3 &=&  \frac{1}{3}[Y,[X,Y]] + \frac{1}{6} [X,[X,Y]], \\
C_4 &=& - \frac{1}{8} ([Y,[Y,[X,Y]]] + [Y,[X,[X,Y]]]) -\frac{1}{24} [X,[X,[X,Y]]]. 
\end{eqnarray*}
Taking into account (\ref{rec.2.3}) and (\ref{rec.2.4}) we have in general 
\begin{equation}   \label{cces.1}
        C_{n+1}  =  \frac{1}{n+1}  \,  f_{[n/2],n} \qquad\quad n \ge 5,
\end{equation}
where the expressions of $f_{n,k}$ are given recursively by  (\ref{rec.2.3}).

In summary, the algorithm we propose for computing the Zassenhaus terms is the following:
\begin{equation}   \label{alg.1}
\begin{array}{l}
  \mbox{Define} \; f_{1,k} \; \mbox{by eq. (\ref{eq:F_1.2})}  \\
   C_n = (1/n) \, f_{1,n-1}, \quad n=2,3,4 \\
   \mbox{Define} \; f_{n,k} \quad n \ge 2, \; k \ge n \;\;  \mbox{by eq. (\ref{rec.2.3})}  \\
   C_n = (1/n) f_{[(n-1)/2],n-1} \quad n \ge 5.
\end{array}
\end{equation}
This constitutes a new recursive way for obtaining directly the term $C_n$ as  a homogeneous
Lie polynomial in $X$, $Y$ of arbitrarily large degree $n$ which can be easily implemented with a symbolic
algebra package.

\subsection{Linear independence}

Algorithm (\ref{alg.1}), or equivalently the procedure given by the identities (\ref{eq:FG})--(\ref{eq:G_n}), provides expressions for $C_n$ that, by construction, involve only 
independent commutators. In other words,
they cannot be simplified further by using the Jacobi
identity and the antisymmetry property of the commutator.

In order to prove this assertion, it is convenient to get a more explicit expression of $F_n(\lambda)$ and
$C_{n+1}$ from (\ref{eq:FG})--(\ref{eq:G_n}). To this end, 
consider for $n\geq 1$ the sets $\J_{n}$ and $\I_{n}$ of $(n+1)$-tuples of non-negative integers recursively defined as follows: 
  \begin{eqnarray*}
    \J_{1} &=& \{(i_0,i_1) \in \N^2\ : \ i_0\geq 1\},\\
    \I_{n} &=& \{(i_0,i_1,\ldots,i_{n}) \in \J_{n} \ : \   i_0 + i_1 + 2 i_2 + \cdots + n i_{n} = n\}, \\ 
    \J_{n+1} &=& (\J_{n}\backslash \I_{n}) \times \N.
  \end{eqnarray*}
The set $\I_{n}$ can be directly defined as the set of $(n+1)$-tuples of non-negative integers satisfying that
$i_0+i_1+ 2 \, i_2 + \cdots+n \, i_{n}=n$ and
\begin{equation}
  \label{eq:I_ncond}
j+1 \leq i_0 + i_1 + 2\, i_2\cdots+ j \, i_{j} \quad \mbox{ for } \quad j=0,\ldots,n-1.
\end{equation}
Thus, in particular, $\I_{1} = \{(1,0)\}$, $\I_2 = \{(1,1,0), (2,0,0)\}$, etc.
Observe that, by construction,  each 
$(i_0,i_1,\ldots,i_n) \in \I_n$ is such that $i_m=0$ if $m>n/2$. 

From (\ref{eq:FG})--(\ref{eq:G_n}), one can then prove by induction on $n$ that, for $n\geq 1$, 
\begin{eqnarray}
\nonumber
  F_{n}(\lambda) &=& \sum_{(i_0,i_1,\ldots,i_n) \in \J_{n}} \frac{(-1)^{i_0+\cdots+i_n} \lambda^{i_0 + i_1 + 2 i_2 + \cdots + n i_{n}}}{i_0!i_1!\cdots i_n!} \ad_{C_n}^{i_n} \cdots \ad_{C_2}^{i_2} \ad_{Y}^{i_1} \ad_{X}^{i_0} Y,\\
  \label{eq:C_n}
C_{n+1} &=& \frac{1}{n+1} \sum_{(i_0,i_1,\ldots,i_n) \in \I_{n}} \frac{(-1)^{i_0+\cdots+i_n}}{i_0!i_1!\cdots i_n!} \ad_{C_n}^{i_n} \cdots \ad_{C_2}^{i_2} \ad_{Y}^{i_1} \ad_{X}^{i_0} Y.  
\end{eqnarray}
In fact, this is clearly true for $n=1$ (equations (\ref{eq:F_1}) and (\ref{c.prim}), respectively), 
whereas successive application of (\ref{eq:FG})--(\ref{eq:G_n}) leads to the general result.

Now, repeated application of  Lazard elimination
principle~\cite{reutenauer93fla}, together with $\I_{1}=\{(1,0)\}$,
$\{C_2\} = \{-\frac12 [X,Y]\}=\{-\frac12 \ad_{Y}^{i_1} \ad_{X}^{i_0}
Y\ : \ (i_0,i_1) \in \I_{1} \}$,
shows that, as a vector space, 
\begin{eqnarray*}
  \mathcal{L}(X,Y) &=& \spn{\{X\}} \oplus \mathcal{L}(\{\ad_{X}^jY\ : \ j\geq0\}) \\
 & = &  \spn{\{X\}} \oplus \spn{\{Y\}} \oplus \mathcal{L}(\{\ad_Y^i \ad_{X}^iY\ : \ i\geq0, j \ge 1\}) \\
&=& \spn{\{X,Y\}} \oplus \mathcal{L}(\{\ad_{Y}^{i_1}\ad_{X}^{i_0} Y\ : \ (i_0,i_1) \in\J_{1}\}) \\
&=& \spn{\{X,Y\}} \oplus \mathcal{L}(
\{ C_2\} \cup \{\ad_{Y}^{i_1}\ad_{X}^{i_0} Y\ : \ (i_0,i_1) \in \J_{1}\backslash\I_{1}\}) \\
&=& \spn{\{X,Y,C_2\}} \oplus \mathcal{L}(\{\ad_{C_2}^{i_2}\ad_{Y}^{i_1}\ad_{X}^{i_0} Y\ : \ (i_0,i_1,i_2) \in \J_{2}\}).
\end{eqnarray*}
More generally, application of Lazard elimination together with
(\ref{eq:C_n}) gives
\begin{eqnarray*}
    \mathcal{L}(X,Y)&\subset& 
\spn{\{X,Y,C_2,\ldots,C_{n}\}} \\
& & \oplus \,  \mathcal{L}(\{\ad_{C_{n}}^{i_{n}}\cdots
\ad_{C_2}^{i_2}\ad_{Y}^{i_1}\ad_{X}^{i_0} Y\ : \
(i_0,\ldots,i_{n}) \in \J_{n}\}) \\
&\subset& 
\spn{\{X,Y,C_2,\ldots,C_{n}\}} \\
& & \oplus \, \mathcal{L}(\{C_{n+1}\}\cup \{\ad_{C_{n}}^{i_{n}}\cdots \ad_{C_2}^{i_2}\ad_{Y}^{i_1}\ad_{X}^{i_0} Y\ : \ (i_0,\ldots,i_{n}) \in \J_{n}\backslash\I_{n}\}) \\
&\subset& 
\spn{\{X,Y,C_2,\ldots,C_{n+1}\}} \\
& & \oplus \,  \mathcal{L}(\{\ad_{C_{n+1}}^{i_{n+1}}\cdots
\ad_{C_2}^{i_2}\ad_{Y}^{i_1}\ad_{X}^{i_0} Y\ : \
(i_0,\ldots,i_{n+1}) \in \J_{n+1}\}).
\end{eqnarray*}

In consequence, 
the terms $\{ \ad_{C_m}^{i_m} \cdots \ad_{C_2}^{i_2} \ad_{Y}^{i_1}
\ad_{X}^{i_0} Y\ : \ (i_0,i_1,\ldots,i_m) \in \J_{n}\}$ are linearly
independent in the free Lie algebra $\mathcal{L}(X,Y)$ and the same is true for the 
representation (\ref{eq:C_n}) of
the Zassenhaus terms.

\subsection{Computational aspects}

We have implemented the recursive procedure (\ref{alg.1})  in \textit{Mathematica}\texttrademark \; as the following
algorithm. 

\begin{alltt}
  Clear[Cmt, ad, ff, cc];
  $RecursionLimit= 1024;
  Cmt[a_, a_]:= 0;
  Cmt[a___, 0, b___]:= 0;
  Cmt[a___, c_ + d_, b___] := Cmt[a, c, b] + Cmt[a, d, b];
  Cmt[a___, n_ c_Cmt, b___]:= n Cmt[a, c, b];
  Cmt[a___, n_ X, b___]:= n Cmt[a, X, b];
  Cmt[a___, n_ Y, b___]:= n Cmt[a, Y, b];
  Cmt /: Format[Cmt[a_, b_]]:= SequenceForm["[", a, ",", b, "]"];

  ad[a_, 0, b_]:= b;
  ad[a_, j_Integer, b_]:= Cmt[a, ad[a, j-1, b]];
  ff[1, k_]:= ff[1, k] = 
     Sum[((-1)^k/(j! (k-j)!)) ad[Y, k-j, ad[X, j, Y]], \{j, 1, k\}];
  cc[2] = (1/2) ff[1, 1];
  ff[p_, k_]:= ff[p, k] = 
     Sum[((-1)^j/j!) ad[cc[p], j, ff[p-1, k - p j]], \{j, 0, 
         IntegerPart[k/p] - 1\}];
  cc[p_Integer]:= cc[p] = 
     Expand[(1/p) ff[IntegerPart[(p-1)/2], p-1]];
\end{alltt}

The object \texttt{Cmt}$[x_1,x_2,\ldots, x_{n-1},x_n]$ refers to the nested commutator  \newline
$[x_1,[x_2, \ldots [x_{n-1},x_n]\cdots ]]$. It has attached just the linearity property (there is no need to attach to it the antisymmetry property and the Jacobi identity). The symbol \texttt{ad} represents the adjoint operator and its powers
$\ad_a^j b$, whereas \texttt{ff[1,k]}, \texttt{ff[p,k]} and \texttt{cc[p]} correspond to expressions 
(\ref{eq:F_1.2}), (\ref{rec.2.3}) and (\ref{cces.1}), respectively. 
Proceeding in this way, we have obtained the
expressions  of $C_n$ up to $n=20$ with  a reasonable computational time and memory requirements.
Thus, for instance, constructing the terms  up to degree
$n=20$ with a personal computer
(2.4 GHz Intel Core 2 Duo processor  with 2 GBytes of RAM) takes less than 20 seconds of CPU time and 35 MBytes of memory. The expression for $C_{20}$ has 48528 terms, all of them independent. The
resulting expressions up to $C_8$ are identical to those expressed in the classical Hall basis.

In Table \ref{tab.1} we collect the CPU time (in seconds) and memory (in MBytes) needed 
to construct the terms $C_2, C_3, \ldots, C_n$ up to a given value of $n$ 
both with the recurrence  (\ref{alg.1}) (\textit{New}) and the implementation provided in
\cite{weyrauch09ctb} using a variant of the so-called comparison method previously introduced in 
\cite{witschel75ooe} (\textit{W-S}).
 Notice
that with this method, which is the most efficient of all the procedures analyzed in \cite{weyrauch09ctb},  the
Zassenhaus exponents $C_j$ are expressed as  linear combinations of words of length $j$ and not directly
in terms of independent commutators (although this is always possible by applying Dynkin--Specht--Wever
theorem or Theorem 2 in \cite{weyrauch09ctb}, and then simplifying the resulting expressions by taking 
into account the Jacobi identity and the antisymmetry property of the commutator). 
For comparison, $C_{16}$ has 54146 terms when expressed
as combinations of words, but only 3711 terms with the new formulation. This translates directly into the memory requirements of both algorithms, as is evident from the results collected in the table.

\

 \begin{table}[thb!]  
 \begin{center}
 \begin{tabular}{|c||c|c||c|c|}
  $n$   &  \multicolumn{2}{c||}{CPU time (seconds)}  &  \multicolumn{2}{c|}{Memory (MBytes)} \\  [0.5ex] \hline
            &  \textit{W-S}       &   \textit{New}     &  \textit{W-S}   &  \textit{New}   \\   [0.5ex] \hline
    14   &  29.18   &   0.14   &   122.90  &      0.88     \\
    16   &  203.85  &  0.59  &  764.32  &        4.09    \\
    18   &                 &    3.01  &                   &        11.12   \\  
    20   &                 &   19.18  &                   &      35.27   \\  \hline
 \end{tabular}
 \end{center}
 \caption{\small{CPU time and memory required for the computation of the Zassenhaus terms $C_2, C_3, \ldots, C_n$
 up to the given value of $n$ using the
 algorithm presented in \cite{weyrauch09ctb} (\textit{W-S}) and recurrence (\ref{alg.1}) (\textit{New}). }}
   \label{tab.1}
 \end{table}

\section{Convergence of the Zassenhaus formula}
 
Suppose now that $X$ and $Y$ are defined in a Banach algebra $\mathcal{A}$, that is to
say, an algebra that is also a complete normed linear space whose norm is submultiplicative, 
\begin{equation}    \label{submult}
    \|X \, Y \| \le \|X\| \, \|Y\|
\end{equation}
for any two elements of $\mathcal{A}$. Notice that for the commutator one has $\|[X,Y]\| \le 2 \, \|X\| \, \|Y\|$. 
Then it makes sense to analyze the convergence
of the Zassenhaus formula  (\ref{zass.1}).  

As stated in the introduction, we are aware of only two previous results establishing sufficient conditions
for convergence
of the form $\|X\| + \|Y\| < r$ with a given $r > 0$. Specifically, Suzuki \cite{suzuki77otc} obtained
$r_s = \log 2 - \frac{1}{2} \approx 0.1931$, whereas Bayen \cite{bayen79otc} proved that the domain
of convergence can be enlarged up to a value of $r_b$ given by the unique positive solution of the
equation
\[
    z^2 \left( 1 + 2 \int_0^z \frac{\e^{2w} -1}{w} dw \right) = 4 (2 \log 2 - 1).
\]    
A numerical computation shows that $r_b = 0.59670569\ldots$. Thus for $\|X\| + \|Y\| < r_b$ one
has
\begin{equation}   \label{conv.1}
   \lim_{n \rightarrow \infty} \e^{X} \, \e^{Y} \, 
   \e^{C_2} \cdots \e^{C_n} = \e^{X+Y}.
\end{equation}

 In the following,  we use recursion (\ref{eq:FG})--(\ref{eq:G_n}) to show that 
(\ref{conv.1}) holds indeed for $(x,y) \equiv (||X||,||Y||) \in \mathbb{R}^2$ in a domain that is larger 
than $\{(x,y) \in \mathbb{R}^2\ : \ 0\leq x+y < r_b \}$.

\begin{figure}[th!]
\begin{center}
%\makebox{\psfig{figure=nf3a1.eps,height=7.2cm,width=8.25cm}}
%\makebox{\psfig{figure=nf3b1.eps,height=7.2cm,width=8.25cm}}
%\scalebox{1}{\includegraphics{domain}}
  \includegraphics[scale=1.0]{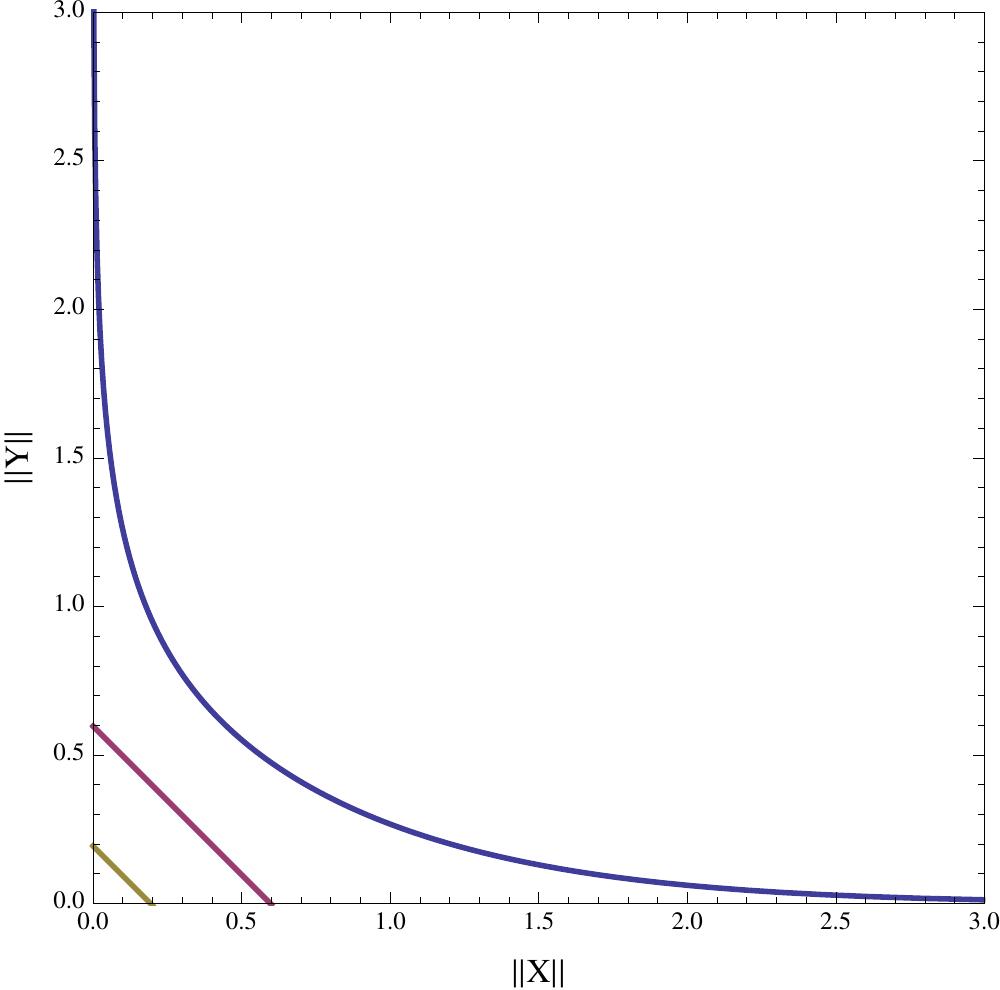}
\end{center}
\caption{{\small Upper boundary of convergence domain for the Zassenhaus formula obtained with the procedure 
(\ref{bounds.1})-(\ref{bounds.2}). Previous results $0 < ||X\|+\|Y\| < 0.1931$ and $0 < \|X\|+\|Y\| < 0.5967$ are 
also depicted for comparison. The new domain contains the region $\|X\| + \|Y\| < 1.054$.}}
\label{fig:1}
\end{figure}

Clearly, (\ref{conv.1}) holds if 
\begin{equation}
  \label{eq:limR}
  \lim_{n\to\infty} ||\R_n(1)||=1,
\end{equation}
where $\R_n(\lambda)$ is given by (\ref{eq:R_n2}), and thus is the solution of the initial value problem
\begin{equation}
\label{eq:ODEaux}
  \frac{d}{d \lambda} R_n(\lambda) = F_n(\lambda) R_n(\lambda), \qquad R_n(0)=I.
\end{equation}
It is well known that, if $\int_0^{1} \|F_n(\lambda)\| d\lambda < \infty$, then there exists a
 unique solution $R_n(\lambda)$ of (\ref{eq:ODEaux}) for $0\leq \lambda \leq 1$, and that  
$\|R_n(1)\| \leq \exp(\int_0^{1} \|F_n(\lambda)\| d\lambda)$. In consequence, 
convergence (\ref{eq:limR})  will be guaranteed whenever  $(x,y)=(\|X\|,\|Y\|) \in \mathbb{R}^2$ 
is such that
\begin{equation*}
\lim_{n\to\infty} \int_0^{1} \|F_n(\lambda)\| d\lambda = 0.
\end{equation*}
From (\ref{eq:C_n}) we have that $\|C_{n+1}\| \leq \delta_{n+1}$, where $\delta_{2}= x \, y$ and for $n\geq 2$,
\begin{eqnarray*}
\delta_{n+1}=\frac{1}{n+1} \sum_{(i_0,i_1,\ldots,i_n) \in \I_{n}} \frac{2^{i_0+\cdots+i_n}}{i_0!i_1!\cdots i_n!} \delta_n^{i_n} \cdots \delta_2^{i_2} y^{i_1} x^{i_0} y. 
\end{eqnarray*}
Similarly, $\|F_{n}(\lambda)\| \leq f_n(\lambda)$, where 
\begin{eqnarray}   \label{f1}
   f_1(\lambda) &=& \sum_{i_1=0}^{\infty}\sum_{i_0=1}^{\infty} \frac{(2 \lambda)^{i_0+i_1}}{i_0!i_1!} 
y^{i_1} x^{i_0} y = \e^{2 \lambda y} (\e^{2 \lambda x}-1) y,
\end{eqnarray}
and for $n\geq 2$,
\begin{eqnarray*}
f_n(\lambda) =
 \sum_{(i_0,i_1,\ldots,i_n) \in \J_{n}} \frac{2^{i_0+\cdots+i_n} \lambda^{i_0 + i_1 + 2 i_2 + \cdots + n i_{n}}}{i_0!i_1!\cdots i_n!} \delta_n^{i_n} \cdots \delta_2^{i_2} y^{i_1} x^{i_0} y.
\end{eqnarray*}
Note that this implies 
\begin{eqnarray*}
  \int_{0}^{1}f_n(\lambda) d\lambda \leq \sum_{k=n}^{\infty} \delta_k,
\end{eqnarray*}
so that (\ref{eq:limR}) is ensured if the series $\sum_{k=2}^{\infty} \delta_k$ converges. Let us analyze each term of
this series by mimicking the recursive procedure given by (\ref{alg.1}). From (\ref{eq:F_1.2}) (or alternatively from
(\ref{f1})) and (\ref{rec.2.3}), we get
\begin{eqnarray}   \label{bounds.1}
   \|f_{1,k}\| & \le &  d_{1,k}  \equiv 2^k y \sum_{j=1}^k \frac{1}{j! (k-j)!} x^j y^{k-j} = \frac{2^k}{k!} y \big( (x+y)^k - y^k \big) 
    \nonumber  \\
  \|f_{n,k}\| & \le & d_{n,k} = \sum_{j=0}^{[k/n]-1}  \frac{2^j}{j!} \delta_n^{j} d_{n-1,k-nj}
\end{eqnarray}
whence
\begin{equation}   \label{bounds.2}
   \|C_n\| \le \delta_n = \frac{1}{n}   d_{[(n-1)/2],n-1}, \qquad n \ge 3.
\end{equation}   
A sufficient condition for convergence is obtained by imposing
\begin{equation}   \label{conv}
   \lim_{n \rightarrow \infty}  \frac{\delta_{n+1}}{\delta_n} < 1.
\end{equation}  
At this point it is worth remarking that, although not reflected by the notation, 
both $d_{n,k}$ and $\delta_n$ depend on $(x,y)=(\|X\|,\|Y\|)$, so condition (\ref{conv}) implies in fact a 
constraint on the convergence domain $(x,y) \in \mathbb{R}^2$ of the Zassenhaus formula. 
In Figure~\ref{fig:1}, we show graphically the (numerically computed) domain $\mathcal{D}$ of such points $(x,y)$.
This has been obtained by computing for each point the coefficients $d_{n,k}$ and $\delta_n$ up to $n=1000$
(in fact, considering a smaller value of $n$ the figure does no change significantly). We have also included
for comparison the previous results $x+y < 0.1931$ and $x+y < 0.5967$ of Suzuki and Bayen, respectively.
Observe that the new convergence domain is considerably larger. In particular,
it contains the region $x+y < 1.054$, but it is not restricted to that. For instance, the convergence domain contains the sets  $\{(x,2.9216)\ : \ x<0.00292\}$ and
$\{(2.893,y)\ : \ y<0.0145\}$, and also the points   $(x,0)$ and $(y,0)$ with arbitrarily large value of $x$ or $y$.

In summary, we have presented a new recursive procedure that not only allows us to get the 
expressions of the
Zassenhaus exponents $C_n$ directly in terms of independent commutators in an efficient way and can be easily implemented in any symbolic algebra package, but also shows, by bounding appropriately each term in the recursion,  that  the convergence domain of the Zassenhaus formula is considerably larger than the domain guaranteed by previously known results.

\subsection*{Acknowledgments}

This work has been partially supported by Ministerio de
Ciencia e Innovaci\'on (Spain) under project MTM2010-18246-C03
(co-financed by FEDER Funds of the European Union) and Fundaci\'o Bancaixa under project
P1.1B2009-55.

\end{document}